\documentclass{article}
\usepackage{times}
\usepackage{geometry}
\usepackage{titlesec}
\usepackage{setspace} 
\usepackage{ragged2e}
\usepackage{hyperref}
\usepackage{graphicx}
\usepackage{caption}
\usepackage{hhline}
\usepackage{fancyhdr}
\usepackage{ifthen}
\usepackage{url}
\usepackage{footmisc}
\usepackage{xcolor}
\usepackage{etoolbox}
\usepackage[style=numeric, sorting=none]{biblatex}
\usepackage{textcase}
\usepackage{tabularx} 
\usepackage{amsmath}
\usepackage{booktabs}

\definecolor{lightgraytext}{gray}{0.4} 

\DeclareCaptionLabelFormat{figlabel}{{FIG. #2. }}
\DeclareCaptionLabelFormat{tablelabel}{{TABLE #2. }}
\titleformat{\section}
  {\fontsize{10}{12}\selectfont\MakeUppercase} 
  {\thesection.} 
  {0.5em} 
  {} 

\titleformat{\subsection}
  {\fontsize{10}{12}\selectfont\bfseries} 
  {\thesubsection.} 
  {0.5em} 
  {} 

\titleformat{\subsubsection}
  {\fontsize{10}{12}\selectfont\itshape} 
  {\thesubsubsection.} 
  {0.5em} 
  {} 

\newcommand{\centeredparheader}[1]{%
    \textcolor{lightgraytext}{%
        \parbox[c]{\textwidth}{%
            \centering #1%
        }%
    }%
}

\fancypagestyle{plain}{
  \fancyhf{}%
  \fancyhead[C]{\centeredparheader{%
    \fontsize{8}{0}{W. A HORNSBY et al.} }}%
}

\newcommand{\evenheader}{\centeredparheader{%
   \fontsize{8}{0}{TH/P1-34} \quad\textcolor{lightgraytext}}}
    
\newcommand{\oddheader}{\centeredparheader{%
    \fontsize{8}{0}{W. A HORNSBY et al.} }}
    
\begin{document}

\title{\raggedright
  \textbf{\fontsize{12}{15}\selectfont
  \MakeUppercase{GAUSSIAN PROCESS SURROGATE MODELS FOR THE PROPERTIES OF MICRO-TEARING MODES IN SPHERICAL TOKAMAKS. \\
  }}
  \vspace{-3.2em}
  \setstretch{0.6}
}
\date{}
\maketitle

\begin{flushleft}
  W.A. HORNSBY, A. GRAY, J. BUCHANAN, D. KENNEDY, B.S. PATEL, F.J. CASSON, C. ROACH\\
  UKAEA \\
  Culham Science Centre, OX14 3DB Abingdon, United Kingdom \\
  Email: william.hornsby@ukaea.uk
\end{flushleft}

\begin{flushleft}
  M. B. LYKKEGAARD, H. NGUYEN, N. PAPADIMAS, B. FOURCIN, J. HART \\
  digiLab Solutions Ltd \\
  The Quay, Exeter. EX2 4AN, United Kingdom \\
\end{flushleft}

\section*{\bfseries\MakeTextUppercase{a}\MakeTextLowercase{bstract}}
{\fontsize{9pt}{12pt}\selectfont\hspace{1cm} Spherical tokamaks (STs) have many desirable features that make them a suitable choice for fusion power plants. To understand their confinement properties, accurate calculation of turbulent micro-instabilities is necessary for tokamak design. Presented is a novel surrogate model for Micro-tearing modes (MTMs), the micro-instability thought to be dominant in high beta STs [1,2,3]. Direct numerical calculation of micro-instabilities is computationally expensive and is a significant bottleneck in integrated plasma modelling. The considerable number of geometric and thermodynamic parameters, the interactions that influence these coefficients and the resolutions needed to accurately resolve these modes, makes direct numerical simulation for parameter space exploration computationally extremely challenging. However, this and the dearth of accurate reduced physics models for MTMs makes it suitable for surrogate modelling using Gaussian Process Regression, a modern machine learning technique [4]. This paper outlines the further development of a data-driven reduced-order model across a spherical tokamak reactor-relevant parameter space utilising Gaussian Process Regression (GPR) and classification; techniques from machine learning.  To build the original simple GP model these two components were used in an active learning loop to maximise the efficiency of data acquisition thus minimising computational cost.  The `simple' GP was seen to show a plateau of fidelity with more data and to be under-confident, particular in areas of parameter space close to marginal stability. It is postulated that the presence of multiple sub-types of MTM could be the root cause, with the underlying function being less smooth than expected.  An expansion of the model using clustering algorithms to find optimal sub models using a mixture of experts approach is shown to greatly improve the variances in the outputs of the GP model. }

\renewcommand{\footnoterule}{\hrule width \linewidth}

\section{Introduction}

Spherical tokamaks hold several potential advantages when compared to conventional aspect ratio devices. Primarily this is the ability to access high $\beta$ regimes of operation, allowing a more compact and ultimately cheaper reactor design. However, in these scenarios electromagnetic instabilities, such as kinetic ballooning modes (KBM) and micro-tearing modes (MTM), are typically excited, both of which can drive radial heat propagation away from the plasma core and degrade plasma performance. 

MTMs exhibit extremely fine radial structure and subsequently require high numerical resolution. Even local linear gyrokinetic simulations are currently prohibitively expensive in the context of integrated modelling.  An alternative to repeated direct use of high-fidelity models in integrated workflows is to create a training data-set in an efficient way, upon which a data-driven reduced-order model of the simulation outputs using machine learning based techniques can be trained.  Gaussian Process (GP) Regression and Classification is a probabilistic, non-parametric Bayesian method [4] for predictive modelling. A GP is specified in terms of its mean function and kernel (covariance function). The kernel expresses the degree of correlation between different locations in the parameter space. Some widely used kernels describe this correlation in terms of correlation length scales; broadly, small (large) length scales allow for small (large)-scale variations. GPs are particularly well suited for data-scarce problems, since they naturally provide both the mean and the variance of the predictions, given sometimes noisy data. 

A GP  based surrogate model of the linear properties of the MTM was constructed using the high-fidelity Gyrokinetic turbulence code, GS2 [5] as the source of the training data  [6].   The ranges of the input parameters considered are described in Table. 1 and represent the prospective limits within which future spherical tokamak power plants are expected to operate [1].   The outputs of the model are the properties of the micro-tearing modes that are important for building a quasi-linear heat transport model: namely the growth rates, frequencies and mode contribution to the radial electron heat flux caused by the interaction of the turbulent field and the turbulent pressure fluctuations.  Further parameters describing the geometry are fixed to set values, described in [6].

\begin{table}[htbp]
\centering
\caption{\raggedright \MakeUppercase {{TABLE SHOWING THE INPUT PARAMETERS WHICH ARE VARIED IN THIS
MODEL WITH THEIR MAXIMUM AND MINIMUM VALUES, ALSO LISTED
ARE THE MODEL OUTPUTS.
}}}
\label{tab:example_table}
\begin{tabular}{l c  c  c }
\hhline{~~~}
\hhline{-|-|-|-}
Variable & Description & Min. value & Max. value \\
\hhline{-|-|-|-}
q & Safety factor & 2 & 9 \\
$\hat{s}$ & Magnetic shear & 0.05 & 5 \\
$a/L_{ne}$ & Normalised electron density gradient & 0 & 10 \\
$a/L_{Te}$ & Normalised electron temperature gradient & 0.5 & 6 \\
$\beta$ & Normalised plasma pressure & 0 & 0.3 \\
$\nu_{ei}$ & Electron-Ion collision frequency [$c_{s}/a$] & 0 & 0.1\\
$k_{y}$ & Binormal mode wavelength [$\rho_{s}$] & 0 &  1\\
\\
Model Outputs & & & \\
\\
$\gamma$ & Mode growth rate [$c_{s}/a$] & & \\
$\omega$ & Mode frequency [$c_{s}/a$] & & \\
$Q_{e}$ & Electron flutter flux normalised to $|A_{||}|^{2}$ & &\\
\\
\hhline{-|-|-|-}
\end{tabular}
\end{table}

The MTMs will be examined using local linear gyrokinetics on a single core flux surface with $r/a = 0.67$, where $a$ is the minor-radius of the last closed flux surface. A Miller parameterisation was used to model the equilibrium. Three kinetic species are modelled: deuterium, tritium, and electrons ($n_{e} = n_{D} + n_{T}$, deuterium and tritium in a 50/50 mix). The simulations are run for a single binormal mode, $k_{y}$. We neglect compressional magnetic field ($B_{||}$) effects to reduce the computational complexity of the model and suppress some types of kinetic ballooning mode which are excited by this component of the magnetic field.  A further set of physics-based filters are used to determine if a simulation is accepted into the training set, namely that the mode frequency is within 50\% of the electron diamagnetic frequency, the mode has the correct tearing parity and that the field line significantly deviates away from its equilibrium flux surface.  

In addition to GS2, the toolchain incorporates Pyrokinetics [7] which is a python library that aims to standardise gyrokinetic analysis by providing a single interface for reading and writing input and output files from different gyrokinetic codes, normalising to a common standard. This enables future interoperability with other gyrokinetic codes.  All variables in this paper utilise the standard normalisations used by Pyrokinetics.

\begin{figure}[htbp!]
\centering
\includegraphics[width=0.9\textwidth]{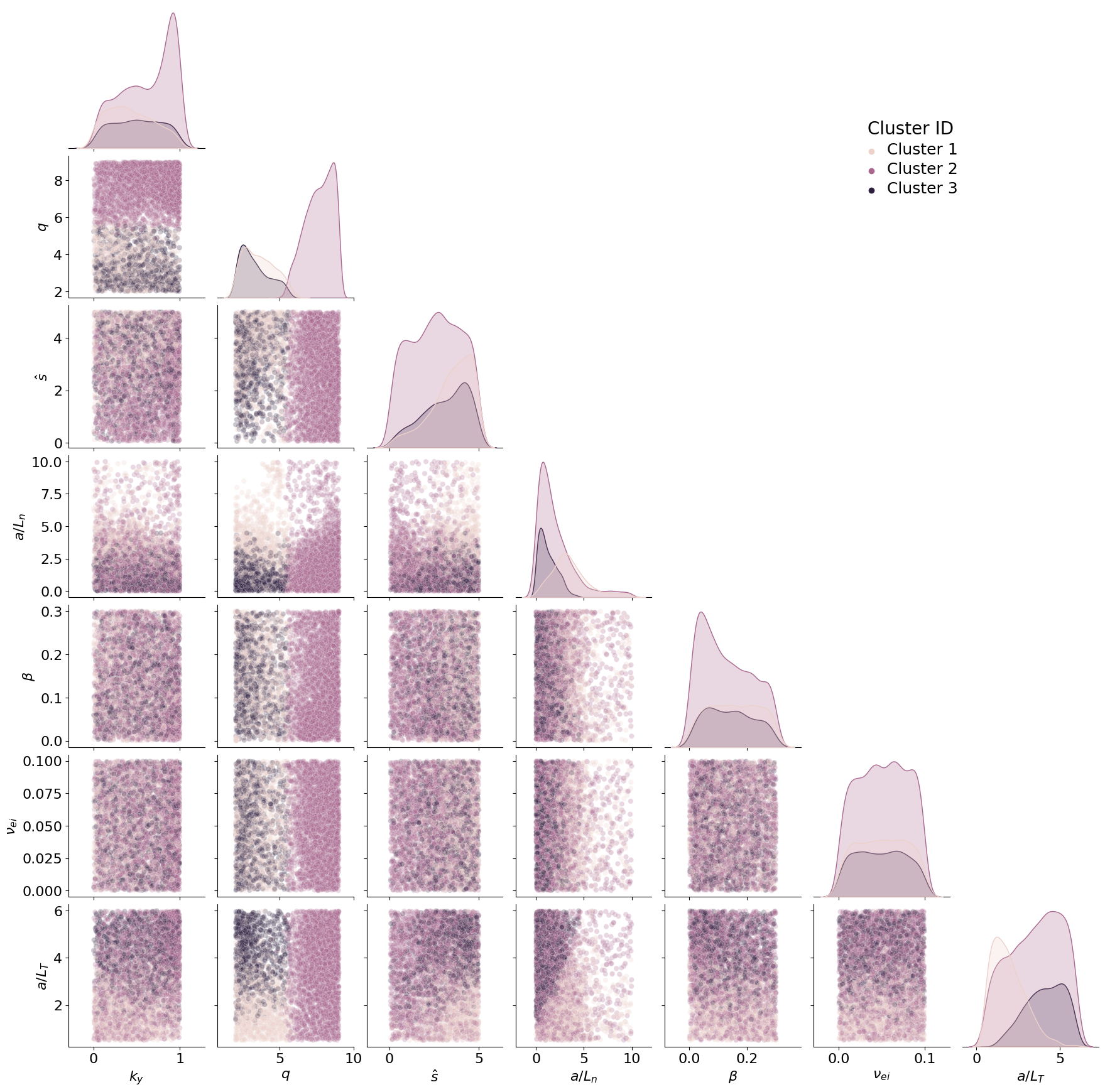}
    \includegraphics[width=0.95\textwidth]{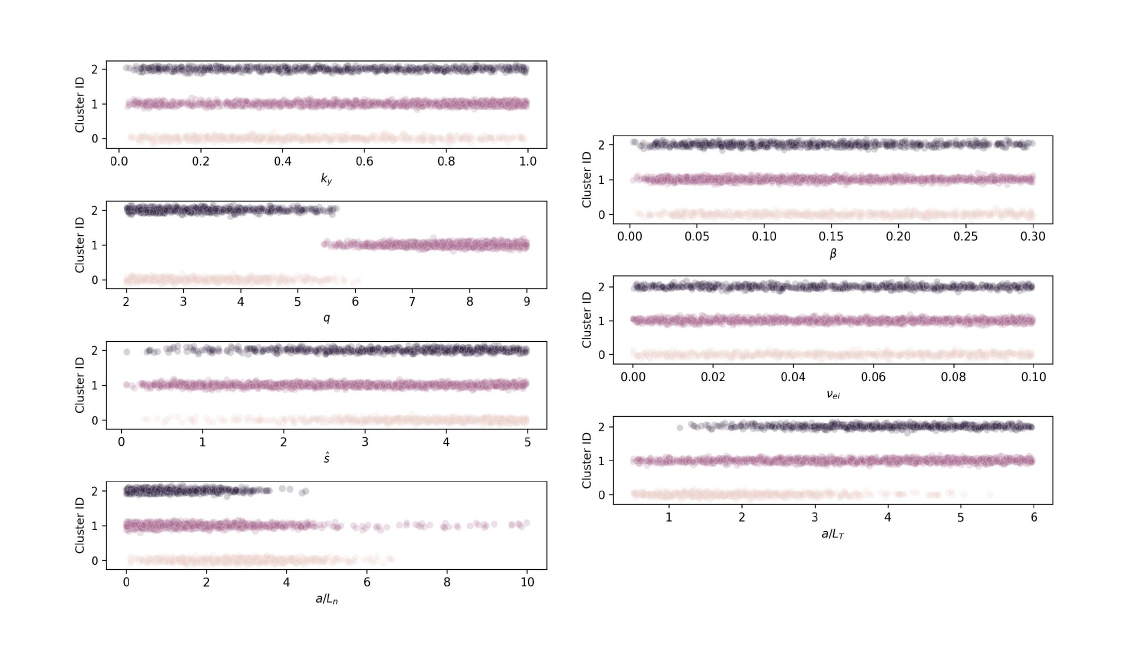}
\captionsetup{font=it}
\caption{(top) The complete MTM data-set with each data point coloured by its respective cluster as determined by our mixture of experts algorithm.  Datapoints that are not determined to be MTMs are not shown.  The algorithm clearly differentiates three separate clusters, a low and high q branch, the former of which is further separated into a low and high temperature gradient branch.  (bottom) Distribution of cluster memberships in the training set over the 7 parameters.}
\label{fig:paramspace}
\end{figure}

\subsection{Classification model}

The classifier is used as a filter for simulation outputs of interest and is trained on a logical parameter indicating whether an MTM is unstable or not (i.e data points that fulfil the criteria described).   This is a binary classification problem, which can be handled using a GP with a Bernoulli distribution as the likelihood and a probit link function, i.e., the Standard Normal Cumulative Density Function (CDF).

\subsection{Regression model}

A regressor models the mode growth rates, frequencies and electron heat fluxes in regions where modes are predicted to exist.  Furthermore, the classifier enables the expansion of the training set by predicting the regions of parameter space where an MTM may be found and thus allowing a targeted use of compute resources. Figure 1 shows the distribution of simulation points found in the database, showing significant fine structure.

Multiple different covariance functions were investigated (RBF and the Matérn family of kernels) for each output variable. Each model used a constant mean function.  There is no appreciable difference between the kernels with respect to the chosen metrics.  None of the kernels tested were significantly better than any others, so we will use the commonly used Matérn 5/2.  

\section{ACTIVE LEARNING PIPELINE}

The global model and active learning pipeline is summarised here.  GS2 is being used as an initial value code, in cases where the simulation is run in areas of stability it will run to a pre-determined maximum time without finding an instability, thus using up significant computational resources without producing useful data upon which to train the regression model. To mitigate this, the classifier is trained to predict the probability that a given point in parameter space will be unstable to MTMs, effectively learning the marginal stability manifold and allowing targeted exploration of the parameter space. The micro-tearing mode is stabilised by some of the parameters that are investigated here, and as such there are large volumes of parameter space where the code may not return a positive result, either a stable mode or a different type of unstable mode that does not fulfil our conditions.  The submanifold characterising marginal stability is of great importance for determining the parameter values beyond which turbulence can form.  

\begin{figure}[htbp]
\centering
\includegraphics[width=0.95\textwidth]{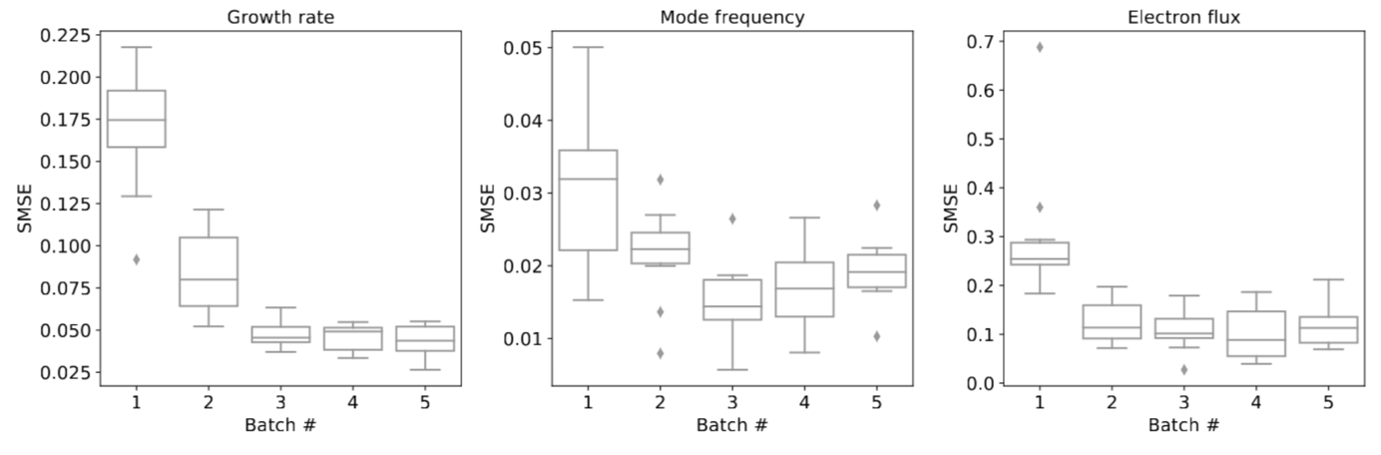}
\captionsetup{font=it}
\caption{The evolution of the standardised mean squared error (SMSE) as a function of the active learning iterations performed, showing that for the growth rate and flux models, a decrease in the error as the data volume is increased is observed. However, adding additional data points in a targeted high flux region of parameter space slightly increased the error in the frequency model. The number of folds is k = 10 for each boxplot.
}
\label{fig:boxplots}
\end{figure}

Initially a small 300-point maxi-min Latin-hypercube sampling is run to fill the parameter space; the classifier and regression model are then built on this data-set. To expand the data-set, the parameter space is sampled using a quasi-Monte Carlo method and the points are passed through the classifier which gives the probability that the parameter set will return an unstable MTM, then finally subjected to rejection sampling in a way that balances exploitation of the current data and exploration of underpopulated areas to prevent biases from the data being reinforced. These samples are then run as batches and incorporated into the complete data-set on which a new regression model is built. This process is repeated until sufficient model fidelity is reached. 

Two general batches of new data were generated using this procedure. The first batch consisted of 1362 accepted samples, of which 971 were subsequently identified as MTMs, corresponding to a hit rate of 71\%. The second batch consisted of 1317 accepted samples, of which 983 were subsequently identified as MTMs, corresponding to a hit rate of 75\%.  This is a significant increase compared to the initial parameter space sampling, which returned only 20\% as unstable MTMs.  This represents a substantial saving in computational time.  The locations of the simulation points in parameter space are shown in figure \ref{fig:paramspace}.  

\begin{figure}[htbp]
\centering
\includegraphics[width=0.95\textwidth]{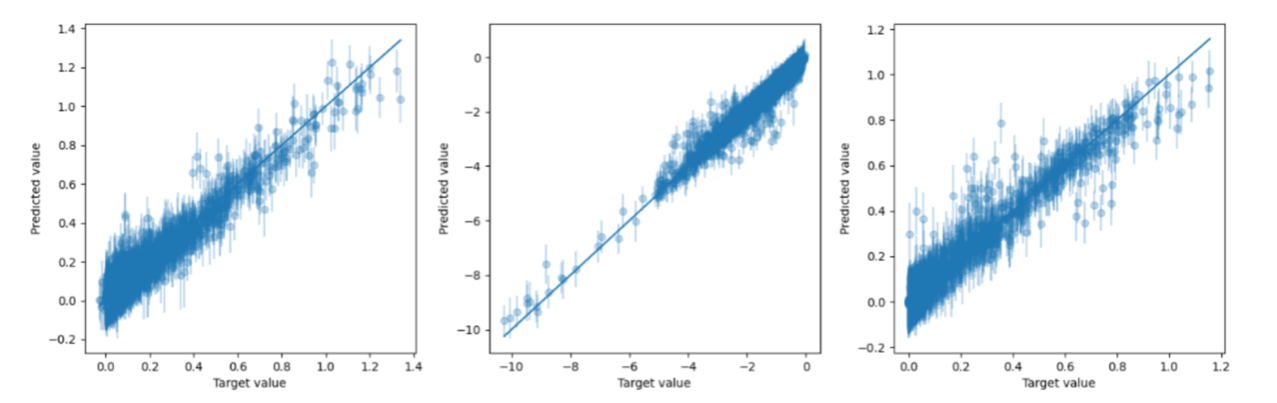}
\captionsetup{font=it}
\caption{Target vs predicted values (error bars display the predictive standard
deviation) for the growth rate (left), mode frequency (centre) and quasilinear flux (right) for the models using the Mat\'{e}rn 1/2 covariance kernel.
}
\label{fig:kfolds}
\end{figure}

The model trained on this expanded data set was observed to lack accuracy at high heat fluxes and so further iterations targeting high flux points were performed. To do this the existing data were labelled according to whether they were below or above a specified threshold and a Bernoulli GP classifier was trained on these labels. Sample points with a high predicted heat flux value were then proposed using the rejection sampling algorithm outlined above, highlighting the flexibility of the classifier driven sampling approach. Using this approach two further iterations targeting $Q_{em,e} > 0.2$ and $Q_{em_e} > 0.5$ were performed.  

Figure \ref{fig:boxplots} shows a box plot for the standardised mean squared error (SMSE) of the three different outputs
as more batches are added to the data-set, showing that for the growth rate and flux models, the predictive loss decreases as the data volume is increased. For the mode frequency models, the variance of the MSLL across batches first increases with batches 2 and 3, and then decreases with batches 4 and 5. Note that batches 2 and 3 were targeted towards areas with a high probability of finding an MTM, while batches 4 and 5 were targeted towards areas with high flux. For each boxplot, the box spans the first quartile (Q1) and the third quartile (Q3), while the centre line displays the second quartile (Q2), i.e., the median value. The whiskers show 1.5 IQR, where IQR is the interquartile range. The points show outliers.  This shows that the active learning pipeline increases the fidelity of the model, while the hit rate of the classifier, which is around 80\% greatly increases the amount of data for a given computational cost.

\section{VALIDATION}

Figure \ref{fig:kfolds} shows the results of a $k$-fold validation of a model trained on the whole data-set and gives an indication of the global performance of the model by comparing the model predicted value against the value as output by GS2.  However, this form of validation does not show how well it can reproduce specific local parametric dependencies, which are important in integrated modelling. 

\begin{figure}[htbp]
\centering
\includegraphics[width=0.95\textwidth]{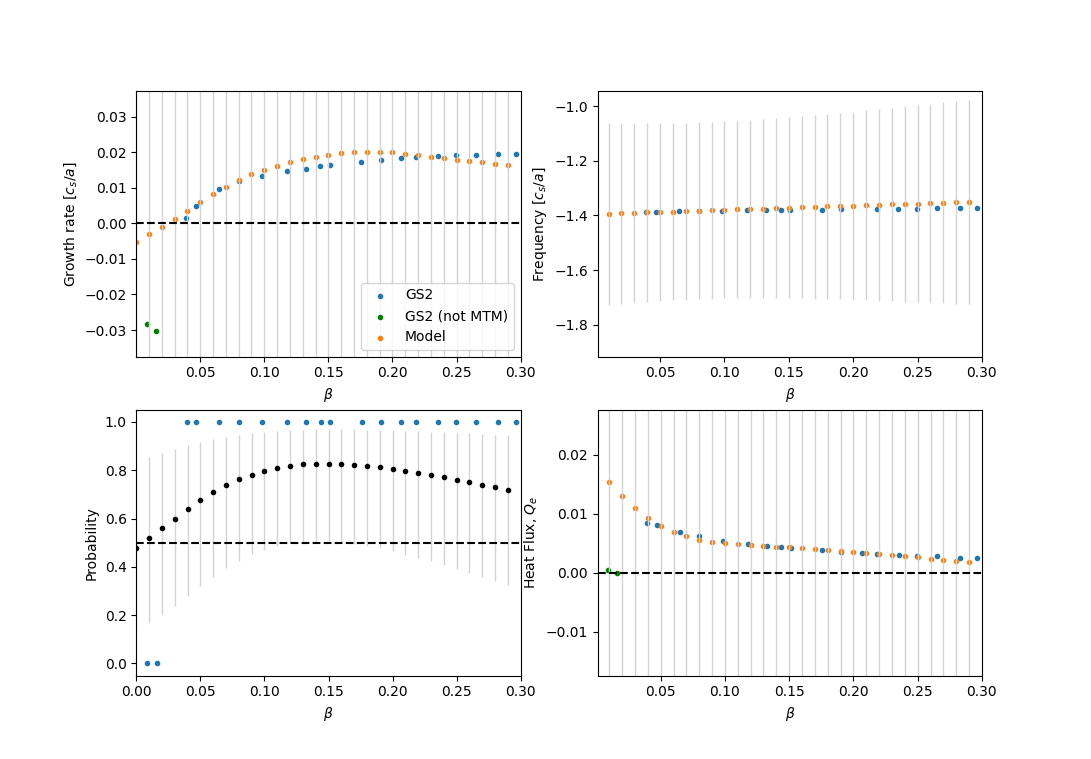}
\captionsetup{font=it}
\caption{A scan over the plasma $\beta$ comparing model output values (orange) with those predicted by GS2 (blue) for (top left) the growth rate, (top right) the mode frequency and (bottom right) the quasi-linear electron flux. The grey bands represent the 95\% confidence interval as predicted by the GP regression. The other model parameters are set to q = 4.45, $\hat{s}$ = 3.0, $k_{y}$ = 0.26, $a/L_{ne}$ = 0.18, $a/L_{Te}$ = 3.86, $\nu_{ei}$ = 0.094. Black dashed lines indicate the zero growth rate line. (bottom left) (black) the classifier predicted MTM probability as a function of $\beta$. (blue) logical parameter denoting a GS2 run with an MTM with positive growth rate.
}
\label{fig:betascan}
\end{figure}

A crucial parameter that shows criticality is $\beta$.  Figure \ref{fig:betascan}. shows a scan in the plasma $\beta$, with all other six parameters fixed (details in figure caption), showing all the features that are encapsulated by the model.  First, it shows that the model captures the stability boundary with high precision, showing an agreement that is significantly better than the error bars suggest, the model also accurately captures the threshold behaviour seen, where both the growth rate prediction and the probability prediction at low $\beta$ drop below zero and to a low probability respectively.  Furthermore, the variation in the mode frequency and the quasilinear flux weight are both captured accurately.  However, it is evident that the model is very under-confident, with large variances seen, particularly in the marginal stability region.   Large uncertainties in the values here would have a significant impact on calculation on the critical gradient needed for turbulence, and greatly impact the margins in reactor design. 

\section{MODE CLUSTERING}

In the global model, it was observed that a plateau in the model fidelity with more data was reached and thus other avenues of investigation would be needed to improve the model fidelity and the confidence intervals, which are of particular importance when determining the location of critical gradients in a transport model.  Standard Gaussian process regression models generally assume a smooth, continuous functional form, making them ideal for common problems in the physical sciences. However in the context of plasma turbulence, theoretical and simulation work has highlighted the existence of relatively sharp discontinuities in functional space due to the presence of multiple MTM mode types, that may hinder the performance of standard Gaussian process models and it was postulated that this was the cause of the plateau with data volume. 

To overcome this limitation, a number of extensions to Gaussian processes have been developed that aim to model the discontinuous nature of some functions. These extensions can be split into two general strategies: 1) deep Gaussian processes [8], and 2) mixture of experts [9]. Deep Gaussian processes make use of a deep hierarchical structure of Gaussian process mappings to learn rich features of the functional space such as discontinuities. On the other hand, mixture of expert models typically use multiple Gaussian process `experts' trained on different subsets of the training data and combine predictions from the experts based in some way on each expert's `expertise' for a given point in input space. Though effective, deep Gaussian processes can be computationally expensive and perform best on large data-sets. For these reasons we chose to use a simple but effective mixture of experts approach in this work. 

Our mixture of experts model was constructed by a three-step process. The first step is to split the data into $k$ subsets. This can be achieved in a variety of ways, but in our case we expect subsets based on geometrical properties of the input space. To generate subsets we employed the $k$-means clustering algorithm as it is inexpensive and fast. Due to the artificial nature of the sampling procedures used to generate our training set, any clustering in the data would be due to artefacts in the data rather than features of the underlying input space. To overcome this problem we included the variables in the output space as inputs to the clustering algorithm (see \autoref{fig:paramspace}). This would prevent prediction of labels to new data points outside the training set, so for the second step we introduce a classification model to predict cluster labels for each data point based only on the input data $X_i$. This prevents train/test leakage, where information leaks from the training set to the test set, affecting the reliability of validation analyses. To predict cluster labels, we used a Dirichlet Gaussian process classifier - a variant of a standard Gaussian process that allows categorical outcome variables by drawing from an underlying Dirichlet distribution. The final step of constructing a mixture of experts models is to fit the expert models to each subset, creating a set of $k$ experts corresponding to the $k$ assumed clusters, $\{e_1, \ldots, e_k\}$. To draw predictions from the mixture of experts for a given data point $X_i$, a weighting $w_k$ of the $k$ experts is required. To obtain these weightings, using the law of total expectation we can draw on the predicted probabilities from the subset classifier $(w_1, \ldots, w_k) \sim \textrm{DirichletClassifier}(X_i)$. We can then use the fact that each expert produces a Gaussian posterior prediction to combine predictions from each expert $(\hat{\mu}_{k}, \hat{\sigma}_{k})$ to a final prediction $\hat{y}_i$ using the following set of equations:
\begin{gather*}
    p(\hat{y}_i | w_{1, \ldots, k}, \hat{\mu}_{1, \ldots, k}, \hat{\sigma}_{1, \ldots, k}) = \frac{1}{\hat{\sigma} \sqrt{2 \pi}} \exp \left[-\frac{1}{2} \left(\frac{\hat{y}_i - \hat{\mu}}{\hat{\sigma}} \right)^2 \right]\\
    \hat{\mu} = \sum_{i = 1}^k w_i \hat{\mu}_i \\
    \hat{\sigma}^2 = \sum_{i = 1}^k w_i \hat{\sigma}_i^2 + \sum_{i = 1}^k w_i \hat{\mu}_i^2 - \left( \sum_{i = 1}^k w_i \hat{\mu}_i \right)^2
\end{gather*}

\begin{table}[]
    \centering
        \caption{\raggedright \MakeUppercase {{Mean squared log loss (MSLL) and standard mean squared error (SMSE) for both the 3-clustered Gaussian process and the standard Gaussian process approaches. In all cases smaller values indicate greater performance. The best performances for each metric and parameter are shown in bold. The clustering approach outperformed the standard approach in all cases except the SMSE for mode frequency, where it performed marginally worse.}}}
    \begin{tabular}{lrrrr}
\toprule
{} & \multicolumn{2}{l}{MSLL} & \multicolumn{2}{l}{SMSE} \\
model & cluster & standard & cluster & standard \\
parameter      &         &          &         &          \\
\midrule
Electron flux, $Q_{e}$   &  {\bf -2.125} &   -1.495 &   {\bf 0.040} &    0.043 \\
Growth rate, $\gamma$    &  {\bf -1.711} &   -1.467 &   {\bf 0.054} &    0.055 \\
Frequency, $\omega$ &  {\bf -2.237} &   -1.656 &   0.034 &    {\bf 0.033} \\
\bottomrule
\end{tabular}

    \label{tab:results}
\end{table}

This approach to discontinuous Gaussian process regression is simple, fast, and easy to implement. It offers several advantages over deep Gaussian processes for small data problems, including improving estimation of noise terms for experts within models, as well as scaling with data-set size much better than standard (or even deep) Gaussian processes. To identify the optimal number of clusters $k$, a simple `elbow' method is used, where the average sum of squared distances to the centroid of each cluster is plotted against number of clusters $k$, and the point where the rate of improvement changes (the elbow) gives the optimal $k$. To verify the performance of our model,  the performance is evaluated for 3 clusters on both a hold-out testing data-set and parameter scans from GS2. 

The complete data-set with each data point coloured by its respective cluster as determined by the mixture of experts algorithm is shown in figure\ref{fig:paramspace}.  The dataset clustering is further shown in the bottom panel of figure\ref{fig:paramspace}, showing the distribution of points within each cluster for each of our seven parameters.  The algorithm clearly differentiates three separate clusters.  Distinct clusters are split along the safety factor $q \approx 5.5$ and to some extent the normalised electron density gradient $a/L_{ne} \approx 2$.

\begin{figure}
    \centering
    \captionsetup{font=it}
    \includegraphics[width=\textwidth]{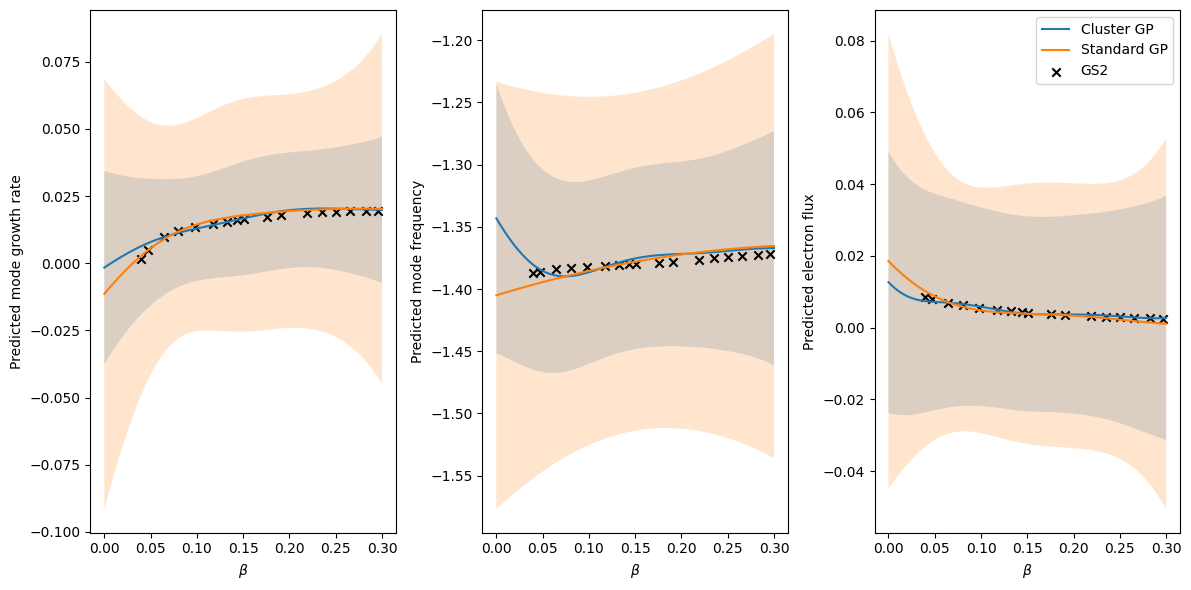}
    \caption{Comparison of predictions made by the cluster and standard GP models against outputs from a GS2 scan over the same input space (See caption to Figure \ref{fig:betascan} for parameters). In all cases the posterior predictive mean of both approaches is accurate, but the cluster GP generally has a narrower prediction interval across the input space.}
    \label{fig:gs2-scan}
\end{figure}

The validation results are shown in \autoref{tab:results} and \autoref{fig:gs2-scan}. It is apparent that the clustered Gaussian process achieved a similar accuracy to the standard Gaussian process but with an improved, narrower, prediction interval. This was reflected in the mean squared log loss (MSLL) scores, which were consistently better for the clustered Gaussian process compared to the standard, whereas the standardised mean squared error (SMSE) was approximately the same for both.

\section{CONCLUSIONS}

This paper described the development of Gaussian process-based models of the linear parameters of micro-tearing modes trained on data generated by a high-fidelity gyrokinetic code. 

The model consists of two components which are combined to form an active learning loop: a Gaussian process classifier which predicts the probability of a given point in parameter space being unstable to MTMs, and a GP regressor which provides the quantity of interest in the unstable regions predicted by the classifier.  Using active learning it was possible to significantly increase the generation efficiency of training data on a previously unexplored parameter space. An accurate model was created using approximately 5000 data points in a 7 dimensional parameter space requiring approximately 1M CPUhrs.

A plateau in fidelity as more data is added has prompted the postulate that this was caused by multiple MTM sub-modes present in the data-set.  Multiple modes with overlapping functions of their growth rates, frequencies and fluxes would mean the underlying function that was being modelled wasn't continuous and smooth.  Standard Gaussian process regression assumes a smooth and continuous function.   To address the discontinuous nature a mixture of experts model was trained on the data-set.

The algorithm clearly finds three distinct clusters upon which GPs are individually trained and then combined.  An improvement in model fidelity is seen, with the MSLL being greatly reduced for all three outputs.


\section*{\hfill \textbf{REFERENCES}\hfill}

\begin{enumerate}
    \renewcommand{\labelenumi}{[\theenumi]}

    \item  WILSON H., CHAPMAN I., DENTON T., MORRIS W., PATEL B., VOSS G., WALDON C., and the Team. STEP — on the pathway to fusion commercialization. 12 2020.

    \item KENNEDY D., GIACOMIN M., CASSON F.J., DICKINSON D., HORNSBY W.A., PATEL B.S., and  ROACH C.M., Electromagnetic gyrokinetic instabilities in the spherical
tokamak for energy production (step) part i: linear physics and sensitivity, 2023.

    \item GIACOMIN M., KENNEDY D., CASSON F.J., AJAY C. J., DICKINSON D., PATEL B.S., and ROACH C.M., Electromagnetic gyrokinetic instabilities in the spherical tokamak for energy production (step) part ii: transport and turbulence, 2023.
    
    \item RASMUSSEN C.E and WILLIAMS C.K.I Gaussian processes for machine learning. Adaptive computation and machine learning. MIT Press, Cambridge, Mass, 2006. 

    \item BARNES M. , DICKINSON D., DORLAND W., HILL  et al. GS2 v8.1.2, 10.5281/zenodo.6882296, 2022.

    \item HORNSBY W.A. et al. {\it submitted}

    \item PATEL B.S, PATTINSON L., HILL P., GIACOMIN M.,KENNEDY D., DICKINSON D., DUDDING H.G., CASSON F.J., and A.C. JAYALEKSHMI A.C. . {\it Pyrokinetics}, 10 2022.

    \item DAMIANOU, A. and LAWRENCE, N.D., "Deep gaussian processes", Artificial intelligence and statistics, Proceedings of Machine Learning Research, 2013

    \item CAO, Y. and FLEET, D.J., Generalized product of experts for automatic and principled fusion of Gaussian process predictions, arXiv preprint arXiv:1410.7827, 2014

\end{enumerate}

\end{document}